\documentclass[%
 reprint,
superscriptaddress,
 amsmath,amssymb,
 aps,
pra,
]{revtex4-1}
\usepackage[dvipsnames]{xcolor}

\usepackage{graphicx}% Include figure files
\usepackage{dcolumn}% Align table columns on decimal point
\usepackage{bm}
\usepackage{hyperref}
\hypersetup{colorlinks = true, linkcolor = [rgb]{0.19411,0.51882,0.667058}, urlcolor = [rgb]{0.125490,0.29542,0.1647058}, citecolor = [rgb]{0.75882,0.37411,0.14117}}

\graphicspath{{../images/}}
\usepackage[toc,page]{appendix}

\newcommand{\ket}[1]{|{#1}\rangle}

\usepackage{braket}

\def\>{\rangle}
\def\<{\langle}

\usepackage[bottom]{footmisc}

\begin{document}
%\fbox{{\scriptsize Preliminary draft.\today}}

\title{Noise-dependent optimal strategies for quantum metrology}

\author{Zixin Huang}

\affiliation{ School of Physics, University of Sydney, NSW 2006, Australia }

\author{Chiara Macchiavello}

\affiliation{Dip. Fisica and INFN Sez. Pavia, University of Pavia, via Bassi 6, I-27100 Pavia, Italy}

\author{Lorenzo Maccone}

\affiliation{Dip. Fisica and INFN Sez. Pavia, University of Pavia, via Bassi 6, I-27100 Pavia, Italy}

\begin{abstract}
We show that for some noisy channels, the optimal entanglement-assisted strategy depends on the noise level. We note that there is a non-trivial crossover between the parallel-entangled strategy and the ancilla-assisted strategy - in the former the probes are all entangled, the latter the probes are entangled with a noiseless ancilla but not amongst themselves. The transition can be explained by the fact that, separable states are more robust against noise and therefore are optimal in the high noise limit, but they are in turn outperformed by ancilla-assisted ones.
\end{abstract}
%\pacs{}
%quantum mechanics, 03.65.Ta
%quantum algorithms and protocols, 03.67.Ac
%quantum optics, 42.50.Gy
%Quantum noise, 42.50.Lc

\date{\today}
\maketitle

\section{Introduction}
Quantum metrology describes strategies which allow the estimation precision to surpass the limit of classical approaches \cite{giovannetti2004quantum,PhysRevLett.96.010401,PhysRevLett.98.090501}. When the system is sampled $N$ times, there are different strategies \cite{rafal} which will allow one to achieve the Heisenberg limit, where the variance of the estimated parameter scales as $1/N^2$. All of these are equivalent when the systems are noiseless. However, in the presence of noise, these strategies are shown to be inequivalent, where entanglement and the use of ancillae are shown to improve the precision of the estimation \cite{guta}.

\begin{figure}[h]
\includegraphics[trim = 1cm 2.5cm 3.5cm 2cm, clip, width=0.9\linewidth]{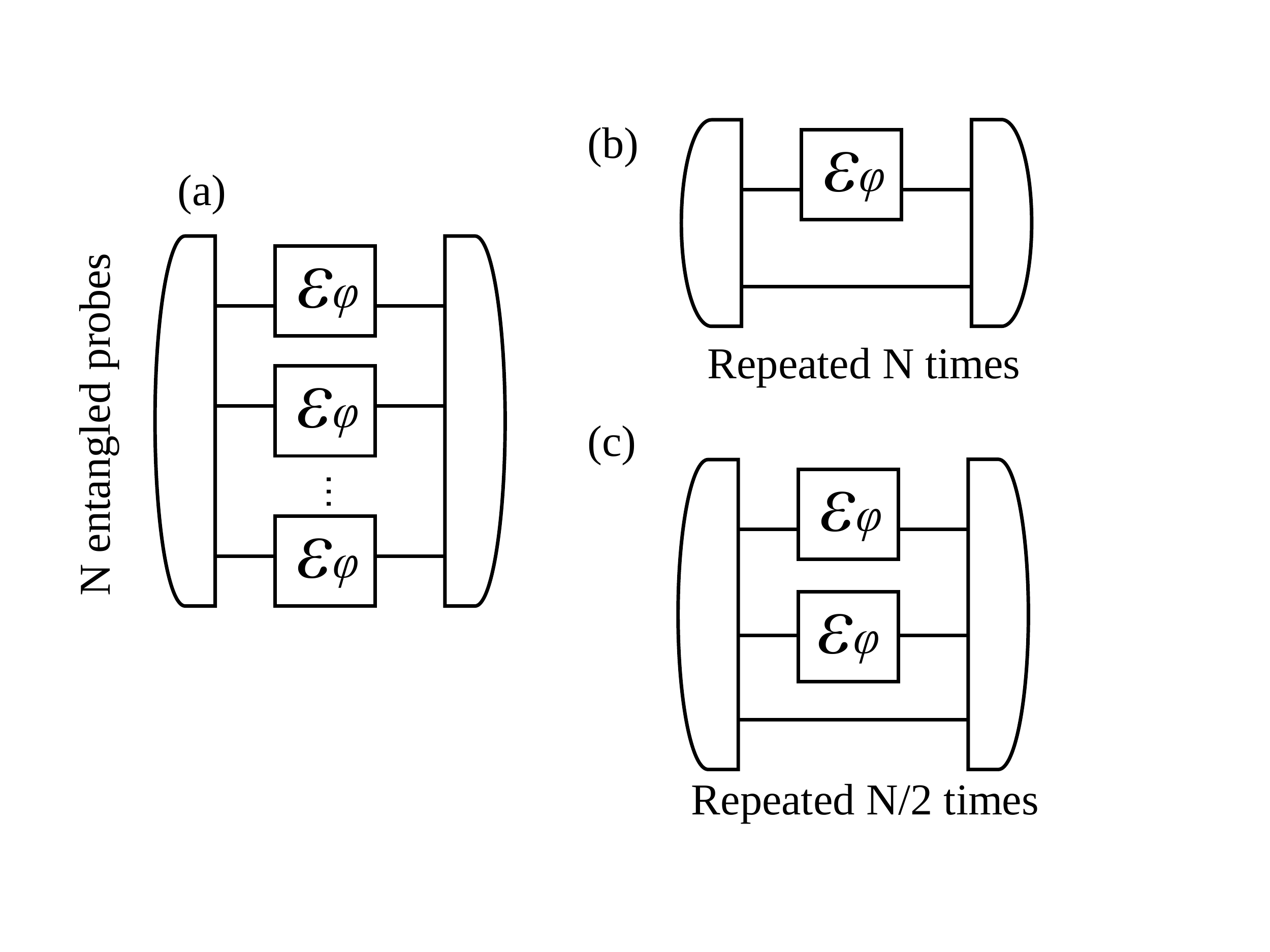}
\caption{\label{f:scheme}
The quantum metrology strategies whose optimality may depend on the noise level of the channel (a) the parallel entangled strategy:  a state of $N$ probes goes through $N$ maps in parallel, this is known to be optimal in the noiseless case. (b) The ancilla-assisted scheme, where $N$ individual probes entangled with a noiseless ancilla go through the map, the probes are not entangled to each other. (c) The intermediate strategy, where two probes are entangled with a noiseless ancilla.}
\end{figure}

In this paper, we show that for some noise channels, the best entanglement-assisted strategy depends on the noise parameter. We note that there is a non-trivial crossover between the parallel-entangled strategy Fig.~\ref{f:scheme} (a) and the ancilla-assisted strategy Fig.~\ref{f:scheme} (b), where the individual probes in the latter case may be entangled with an ancilla but not to each other.  
One would expect the performance of the ancilla-deficient strategy (Fig.~\ref{f:scheme} (c)) to lie in between that of (a) and (b), which we show to be true.

In optical interferometry, in the noiseless regime, $N$ maximally entangled probes in a NOON state \cite{Dowling2008} promises to provide Heisenberg limited sensitivity. However, in the presence of noise, the NOON state is no longer optimal, since losing information on a single probe renders the entire state useless.
In the presence of loss, NOON states can be made more robust (at the expense of reduced sensitivity) by populating additional components in the probes' Hilbert space \cite{prl102040403,dinani}. As the noise parameter increases, the optimal probe becomes less entangled.

Performance of the parallel-entangled scheme in the presence of noise has been extensively investigated (see for example Ref.~\cite{kolodynski2013efficient,chiara,guta,davidovich}), it has been shown that the quantum enhancement in the asymptotic limit of large $N$ amounts to a constant factor \cite{guta,davidovich}. On the other hand, unentangled probes are typically less noise-sensitive and are therefore are optimal in the high noise regime.

% 
% In some cases, this includes ancilla-assisted probes, where additionally no extra physical qubit is required for the ancilla \cite{PhysRevA.94.012101}. 
A strategy to reduce the effect of noise is to use an ancillary system that is entangled with the probes but does not participate in the estimation \cite{rafal}.
% 
% The ancilla-assisted strategy employs pre-shared entanglement between the ancilla and the probe, where the ancilla does not interact with the system being sampled. 
% 
It has been shown for many channels that the ancilla is useful for all levels of the noise parameter \cite{PhysRevA.94.012101,kolodynski2013efficient}. (In bosonic loss channels, ancillae do not provide an advantage in the small $N$ limit that we examine. See for example Ref.~\cite{PhysRevA.90.023856} for engineering states that are more tolerant to loss.) Since unentangled probes perform better than entangled ones in the high noise regime, which is in turn out-performed by ancilla-assisted ones, when comparing the unentangled ancilla-assisted strategy to the parallel-entangled strategy, there must be a crossover in the performance of the strategy depending on the noise parameter: we show there is a large class of noise channels where this transition occurs. 
% Moreover, it is well known that for single parameter estimation, the phase variance lower bounded by the quantum Fisher information is always achievable \cite{matsumoto2002new} since they are constructed from state optimization.

Here we focus on Markovian phase-covariant noise channels \cite{PhysRevLett.116.120801} where the noise map commutes with the parameter encoding.
We find the optimal state for each of the strategies either analytically or numerically, and compare their quantum Fisher information (QFI). We see that this crossover occurs in channels with noise types including amplitude damping (first noted in Ref.~\cite{rafal}), relaxation-excitation (Pauli x and y noise), isotropic depolarization, and those which are a combination thereof.

% Given any Kraus operators, the QFI for the ancilla-assisted state is upper bounded by \cite{guta,rafal}

% \begin{equation}
% F \leq 4 \text{min}_{K} \{  ||\alpha_K||     \}, \alpha_K = \sum_{i} \dot K_i^\dagger \dot K_i (\phi)
% \end{equation}

The results we obtain are based on the quantum Cramer-Rao (QCR) bound \cite{holevo,helstrom,caves,caves1}. It is a lower bound to the
variance of the estimation of a parameter $\varphi$ encoded onto a state $\rho_\varphi$.
For unbiased estimators, $\Delta\varphi ^2 \geqslant 1/\nu
  J(\rho_\varphi)$, where $\nu$ is the number of times the estimation
is repeated, and $J$ is the quantum Fisher information (QFI)
associated with the global state $\rho_\varphi$ of probes and ancillae
(after the interaction $\mathcal{E}_\varphi$ with the probed system). When there is a unique most probable estimate, the bound is achievable in the asymptotic limit that $\nu \rightarrow \infty$.  The QFI is

\begin{eqnarray}
J(\rho_\varphi)=\sum_{j,k:\lambda_j+\lambda_k\neq 0}
{2|\<j|\rho'_\varphi|k\>|^2}/({\lambda_j+\lambda_k})
\label{eq:QFI}\;,
\end{eqnarray}
where $\rho'_\varphi=\partial\rho_\varphi/\partial\varphi$,
$\lambda_j$ and $|j\>$ are the eigenvalues and eigenvectors of
$\rho_\varphi$. The map $\mathcal{E}_\varphi$ encodes the phase parameter
$\varphi$ onto the probes: $\rho_\varphi=\mathcal{E}_\varphi[\rho]$, $\rho$ being 
the initial state. For qubit
channels we suppose that the phase is encoded onto the computational
basis by the unitary $U_\varphi=|0\>\<0|+e^{i\varphi}|1\>\<1|$.

The structure of the paper follows. In Sec.~\ref{sec:pauli} we examine the noise channels describable by a general Pauli channel. 
In Sec.~\ref{sec:adc} we investigate the amplitude damping channel. In the Appendix \ref{app:a} we detail the two-qubit case for a general phase covariant noise model and show that the ancilla provides an advantage over a large range of parameters in state space. In Appendix.~\ref{app:b} we extend our studies to noise parameter estimation, where different quantum metrology strategies are applied to estimating the depolarizing probability of the depolarizing channel.
% \color{blue}
In the subsequent text, we denote the term ``ancilla-assisted strategy" to the optimal two-qubit strategy where one is the ancilla and the other is the probe, and ``intermediate strategy" to mean where there are a larger number of probes than ancillae.

\section{Pauli noise channels \label{sec:pauli}}

% We start by analysing the unital channels generalizable by the Pauli channel, which is described by
% \begin{equation}
% \mathcal{E}[\rho]=  (1- {p_1 - p_2 -p_3}) \rho + {p_1} \sigma_x \rho \sigma_x ^\dagger 
%  + {p_2 }\sigma_y \rho \sigma_y ^\dagger +  {p_3 }\sigma_z
%  \rho \sigma_z ^\dagger, 
% \label{eq:pauli}
% \end{equation}
% where $\sigma_x$, $\sigma_y$, $\sigma_z$ are the Pauli matrices, and $p_1$,$p_2$, $p_3$ are their respective probability of occurring. 
% The Pauli noise channel is phase-covariant when $p_1=p_2$. We analyze the two special cases when $p_3=0$ (excitation/relaxation), and when $p_1=p_2=p_3$ (isotropic depolarization).

\subsection{Pauli x-y noise}
We start by analysing the Pauli channel where the $\sigma_x$ and $\sigma_y$ noise occur with the same probability $\eta/2$.  
% We show that the crossover occurs for $N=2$ probes at efficiency $1-\eta \approx 0.65$.
% , and asymptotically for large $N$ at efficiency $1-\eta \approx 0.61$. 
The channel is described by
\begin{equation}
\mathcal{E} [\rho] \rightarrow (1-\eta) \rho + \frac{\eta}{2} (\sigma_x \rho \sigma_x^\dagger + \sigma_y \rho \sigma_y ^ \dagger),
\label{eq:xy}
\end{equation}
\noindent where the efficiency $(1-\eta)$ is the probability that the transmission is noiseless.
For this channel, the optimal ancilla-assisted state is $1/\sqrt{2}(\ket{00}+\ket{11})$, which has QFI $(1-\eta)$. Since the noise channel is unital, one would expect the optimal state to be symmetric: this is indeed the case. When noise acts upon the probe state, it populates the $\ket{01}$ and $\ket{10}$ components, which is orthogonal to the probe state subspace and therefore can be distinguished.

For $N=2$, in the interval \mbox{$(1-\eta) \in [0.47,1]$}, the optimal parallel-entangled state is $1/\sqrt{2}(\ket{00}+\ket{11})$, 
whose QFI is $4 (\eta-1)^4/(2 \eta^2-2 \eta+1)$. The cross-over between the two strategies occurs at $(1-\eta)\approx 0.647$.

\begin{figure}[h]
\includegraphics[trim = 0 0 0 0, clip, width=0.9\linewidth]{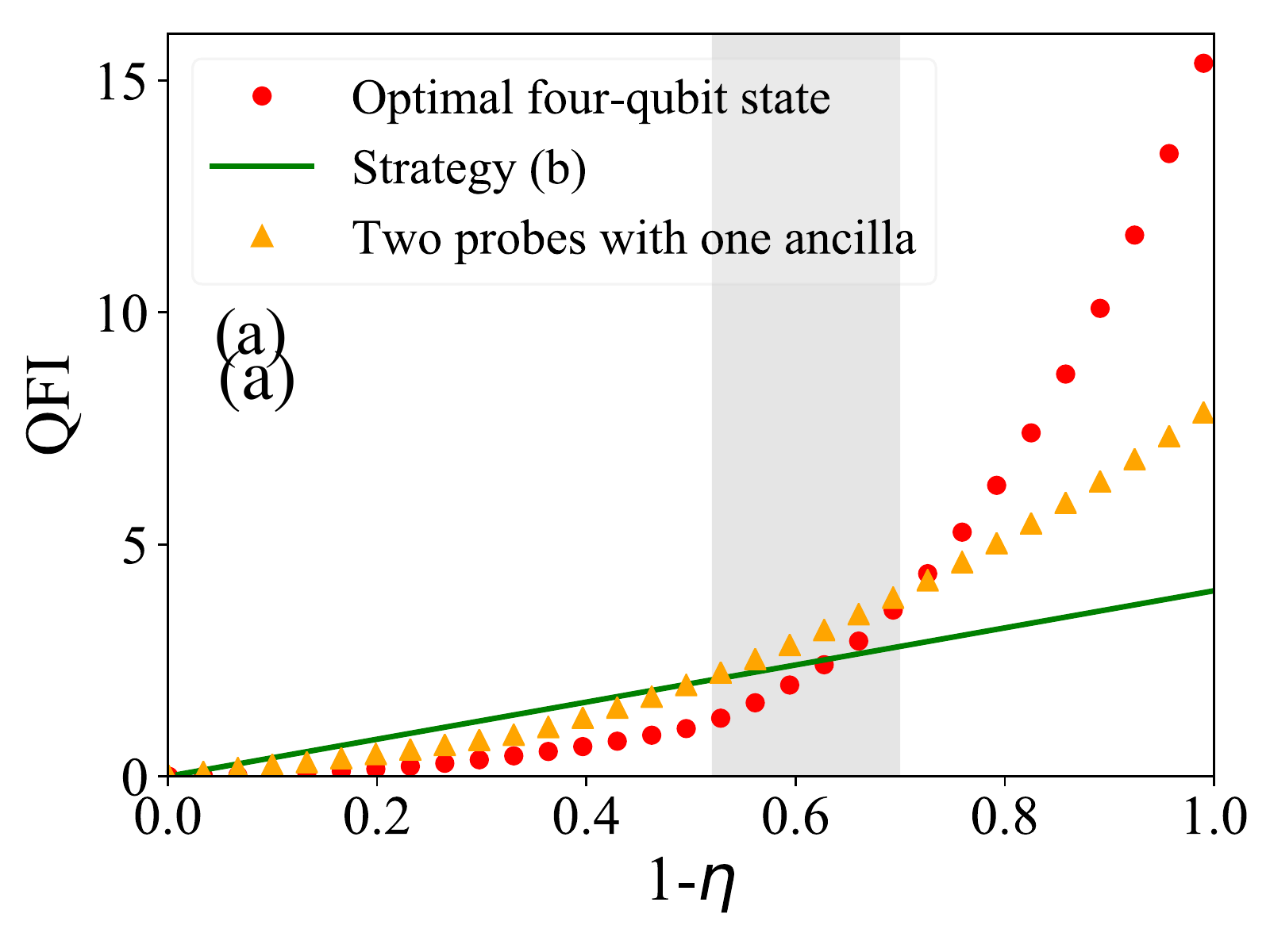} 
\includegraphics[trim = 0 0 0 0, clip, width=0.9\linewidth]{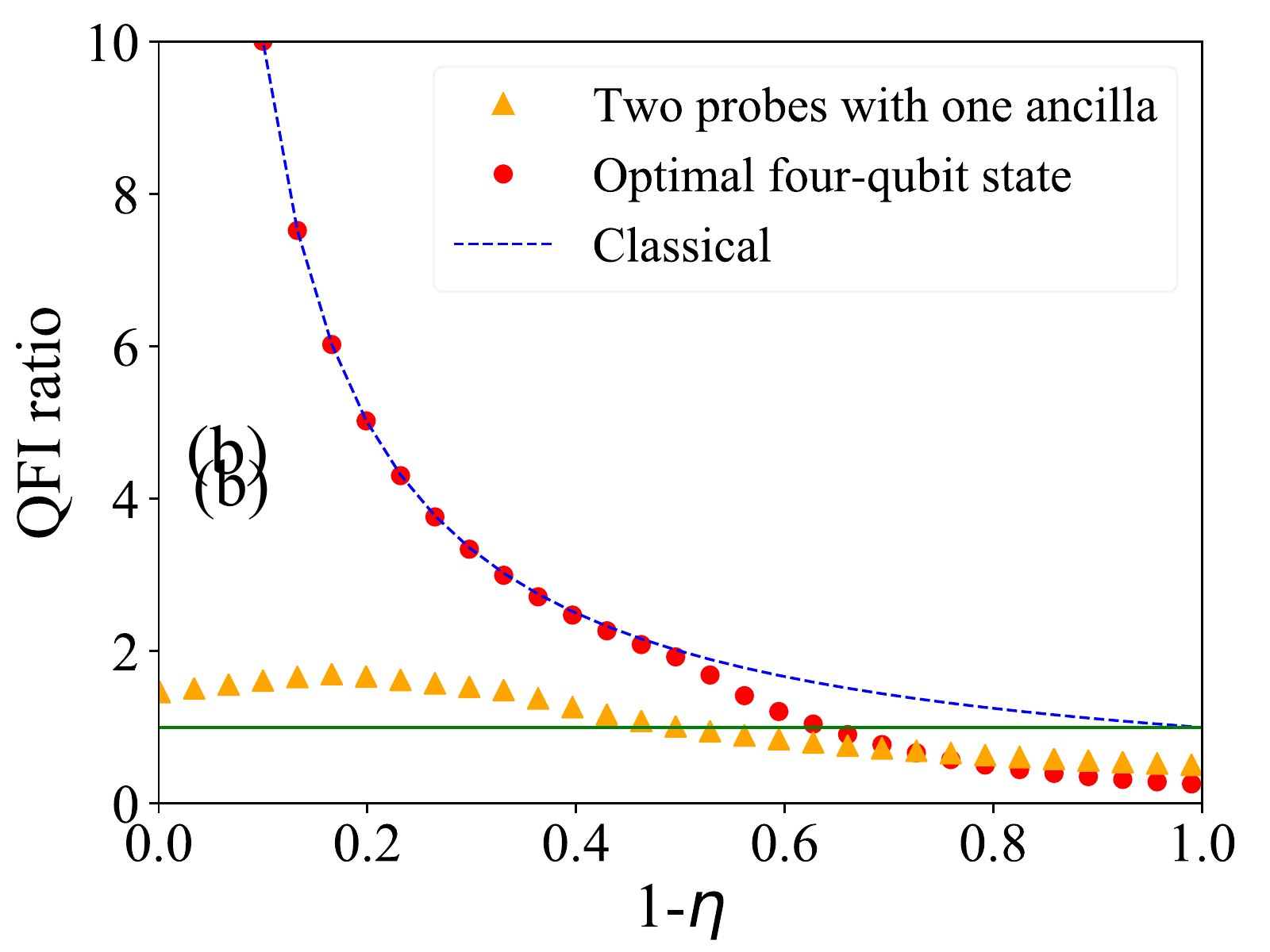}
 \vspace{-3mm}
\caption{\label{f:crossover_xy} Pauli x-y noise: (a) QFI for the ancilla-assisted state $\frac{1}{\sqrt{2}}(\ket{00}+\ket{11})$ where one qubit is the probe and the other is an ancilla (green solid line) used 4 times, the optimal parallel-entangled four-qubit state (red circles), the optimal three-qubit state where two are probes and one is an ancilla used twice (orange triangles) and the upper bound for ancilla-less strategy (blue dashed line) in the presence of Pauli $x$ and $y$ noise occurring with equal probability. The shaded region denotes where the ancilla-deficient state is optimal. (b) The QFI ratio of the ancilla-assisted strategy to the optimal four-qubit state (red circles), the ancilla-assisted strategy to the intermediate state (yellow triangles), and the ancilla-assisted strategy to the best classical strategy (blue dashed line). Since the QFI of the ancilla-assisted strategy is concave as $\eta \rightarrow 1$ and the other two are convex, the advantage becomes increasingly obvious in this regime.}
\end{figure}

To illustrate the difference between the three strategies in Fig.~\ref{f:scheme}, we show the QFI of the optimal states in Fig.~\ref{f:crossover_xy} for $N=4$.

Here we observe two transitions: the ancilla-assisted strategy is optimal in the interval \mbox{$(1-\eta )\in [0,0.52]$}, the intermediate strategy in \mbox{$ [0.52,0.7]$} and the parallel-entangled strategy in \mbox{$ [0.7,1.0]$}. As expected, the performance of the intermediate strategy is in between the other two.

This can be explained by the fact that in the low noise regime, QFI of the parallel-entangled strategy approaches $N^2$ and the ancilla-assisted strategy approaches $N$. Therefore as the noise parameter $\eta\rightarrow 0$, the parallel-entangled strategy is optimal. As the noise parameter increases, the optimal 4-qubit state becomes less entangled, and in the limit that $\eta \approx 1$, the optimal state is $1/\sqrt{2}(\ket{0}+\ket{1})^{\otimes 4}$ and has QFI $4(1-\eta)^2$. In this regime, the ancilla-assisted state whose QFI is $4(1-\eta)$ performs better. Therefore there must be a transition point at which their performance crosses over.

In reality, one may be more 
interested in knowing a strategy beats the best classical strategy 
even in the presence of large noise.
The best classical strategy, i.e. one that only 
uses qubits in the optimal initial state without entanglement or 
ancillae are to use the $\ket{+}$ state, which has QFI $(1-\eta)^{2}$. 
Therefore, for $N$ uses of the channel,the QFI is
\begin{equation}
J(\mathcal{E}[\rho]^{\otimes N}) = N  (1-\eta)^{2}.
\label{eq:adcclassical}
\end{equation}
% \color{blue}
The QFI ratios of the ancilla-assisted state to \eqref{eq:adcclassical}, to the intermediate strategy and to the optimal 4-qubit state are shown in Fig.~\ref{f:crossover_xy} (b). Values above 1 mean that the ancilla-assisted strategy is superior. It is interesting to note that for the Pauli x-y noise channel, the ratio of the ancilla-assisted case to the ancilla-less case becomes increasingly significant as the noise parameter increases (in fact, the ratio approaches infinity as $\eta$ approaches 1). 

\color{black}

\subsection{Depolarizing noise channel}
Next we turn to the depolarizing channel, where we observe the transition numerically for small $N$. The channel is given by
 \begin{eqnarray}
\mathcal{E}[\rho]&=& (1-\eta) \rho + \eta \frac{\openone}{d} \nonumber \\
                 &=& (1- \frac{3 \eta}{4}) \rho + \frac{\eta}{4}( \sigma_x \rho \sigma_x ^\dagger 
 + \sigma_y \rho \sigma_y ^\dagger + \sigma_z  \rho \sigma_z ^\dagger).
\label{eq:pauli}
\end{eqnarray}

\noindent where the efficiency $(1-\eta)$ is the probability that the transmission is noiseless, and $d$ is the dimension of the system the noise acts upon. 
For the ancilla-assisted strategy, the optimal ancilla-assisted state is 
\mbox{${1}/{\sqrt{2}}(\ket{00}+\ket{11})$}, which has QFI \mbox{$2 (1-\eta))^2/(2-\eta)$}.

% \color{blue}
For any $N$, it was shown in Ref.~\cite{1402.0495} that the QFI of the parallel-entangled strategy is upper bounded by 
\mbox{$N/(e^{\gamma_0+\gamma_1+\gamma_2} -1)$} where \mbox{$\gamma_0,\gamma_1,\gamma_2$} are the noise parameters of pure dephasing, excitation and relaxation respectively \cite{1402.0495}.
The depolarizing channel can be rewritten in terms of these parameters, where $2\gamma_0=\gamma_1=\gamma_2=\gamma$. For the depolarizing channel, $\eta \equiv 1-e^{-2\gamma}$, which gives an upper bound of 
$N/(e^{5\gamma/2} -1)= N (1-\eta)^{5/4}/(1-(1-\eta)^{5/4})$. The bound is valid and tight in the high noise regime, where $N(e^{\gamma_0+\gamma_1+\gamma_2}-1) \gg 1$ \cite{1402.0495}.
This upper bound is larger than that of the ancilla-assisted state. For small $N$, we observe the crossover numerically. (The analysis for the bounds is not applicable to the Pauli x-y noise channel, and is therefore omitted in the previous subsection.)

\color{black}

For \mbox{$N=2$}, for the parallel-entangled strategy the transition occurs at $1- \eta \approx 0.65$. In the high noise regime, the optimal state takes the form \mbox{$\epsilon/\sqrt2(\ket{00}+\ket{11}) + \sqrt{1-\epsilon^2}/\sqrt 2(\ket{01}+\ket{10})$}, $\epsilon$ is some parameter that varies with $\eta$, where in the high noise regime, $\epsilon\rightarrow 1/\sqrt 2$. As $\eta$ decreases, the optimal probe state becomes more entangled, and the maximally entangled state is optimal in the interval $(1-\eta) \in [ 0.82,1]$.

% In Fig.~\ref{f:twodepol} we show for $N=4$ the QFI of the three strategies. The results shown are: the optimal parallel-entangled 4-qubit state (red circles), the ancilla-deficient state, where within a 3-qubit state, two are probes and one is an ancilla (repeated twice, orange triangles) and the ancilla-assisted state (repeated 4 times, solid green line), and the asymptotic bound (blue dashed line). 

\begin{figure}[hbt]
\includegraphics[trim = 0 0 0 0, clip, width=0.9\linewidth]{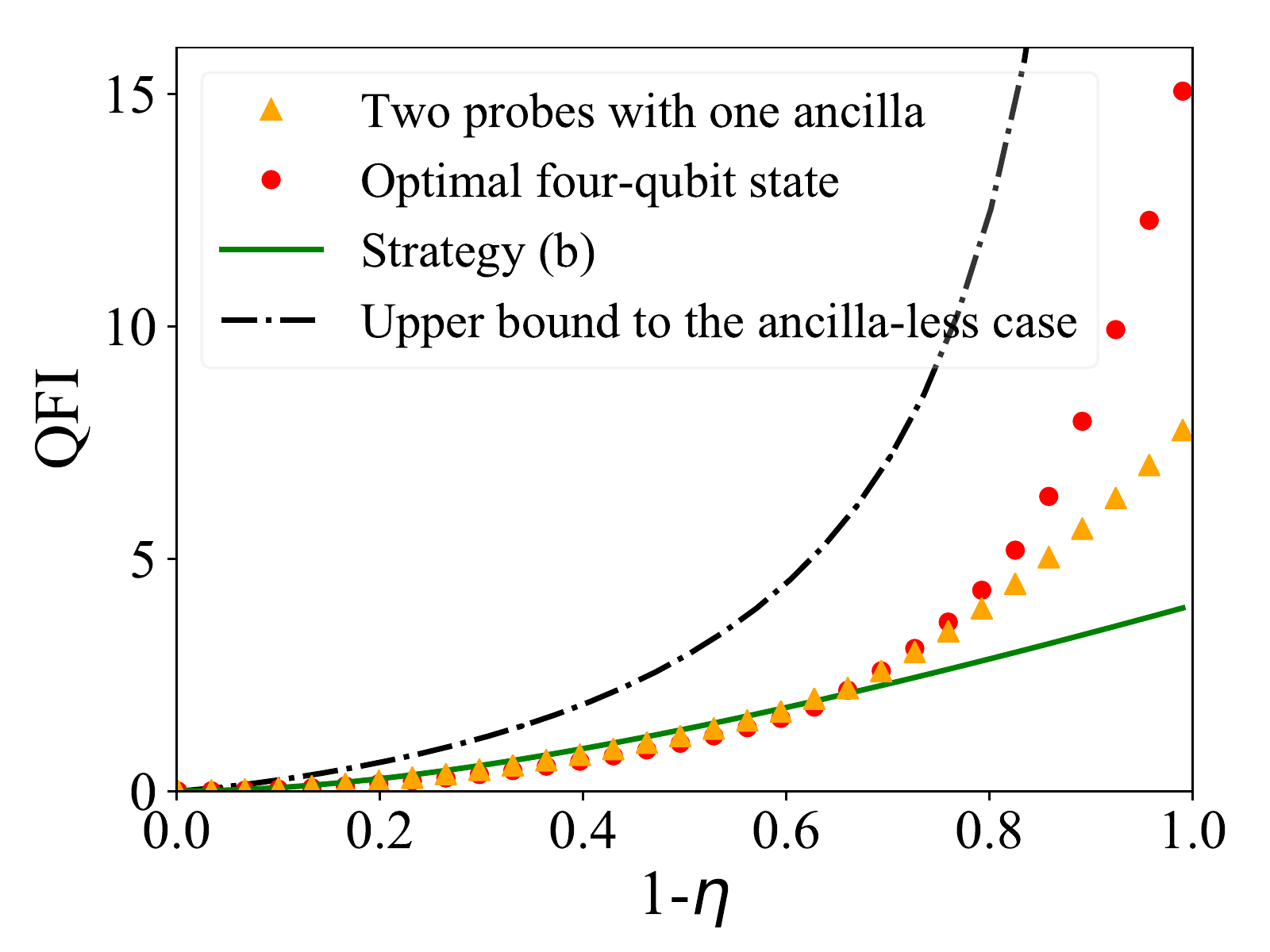}
\includegraphics[trim = 0 0 0 0, clip, width=0.9\linewidth]{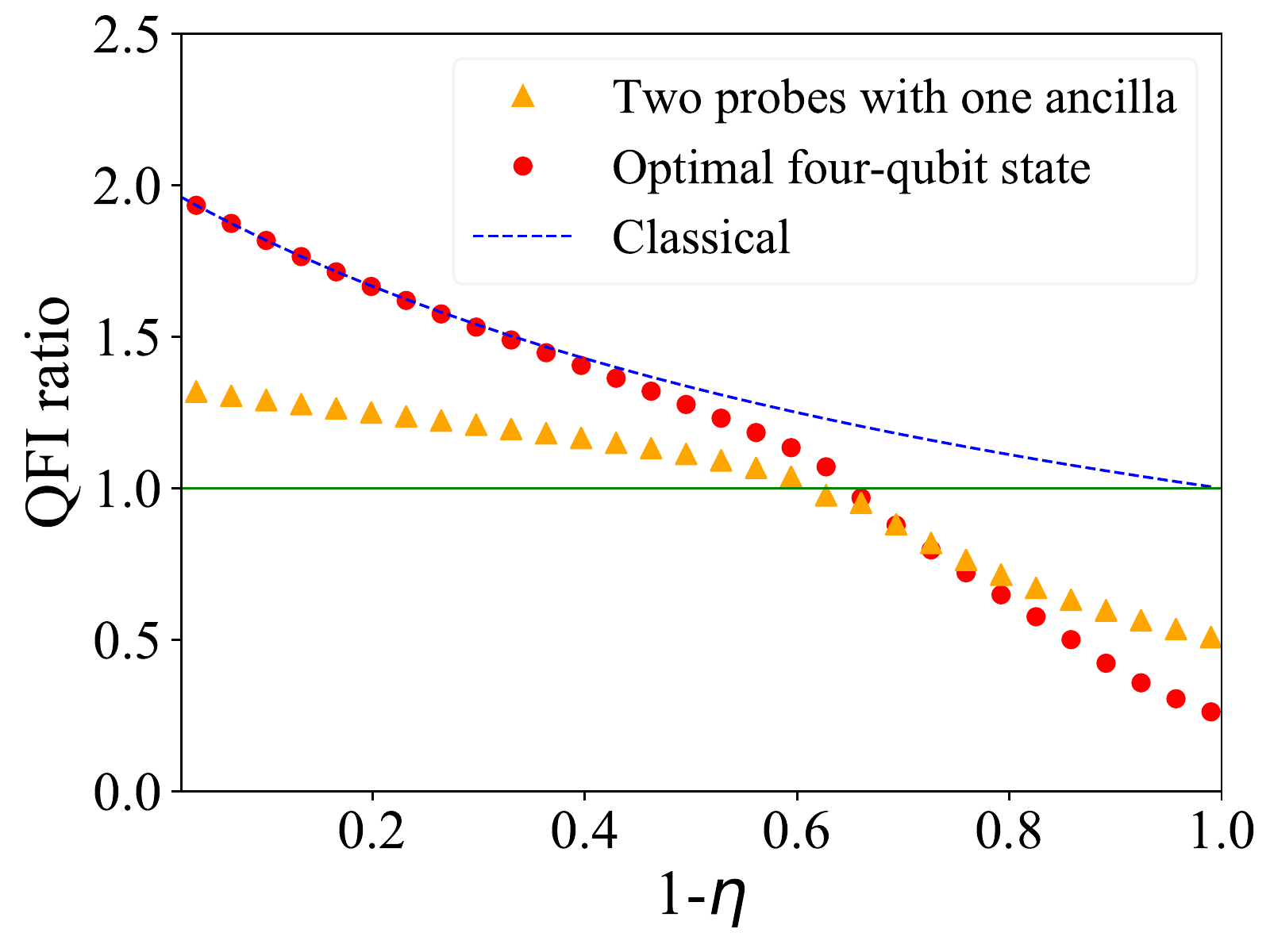}
\caption{ \label{f:twodepol} Depolarizing noise: (a) QFI for the ancilla-assisted state  $\frac{1}{\sqrt{2}}(\ket{00}+\ket{11})$ where one is the probe and the other is an ancilla used 4 times (green solid line), the optimal parallel-entangled four-qubit state (red circles), the optimal three-qubit state where two are probes and one is an ancilla used twice (orange triangles) and the upper bound to the ancilla-less case (black dashed line) in the presence depolarizing noise. 
(b) The QFI ratio of the ancilla-assisted strategy to the optimal four-qubit state (red circles), the ancilla-assisted strategy to the intermediate state (yellow triangles), and the ancilla-assisted strategy to the best classical strategy (blue dashed line).}
\end{figure}

For $N=4$, when $(1-\eta)$ is less than $\approx 0.65$, the ancilla-assisted strategy performs better than the parallel-entangled strategy (and vice versa), see Fig.~\ref{f:twodepol}.
It is interesting to note that, unlike in the channels with Pauli x-y noise and the amplitude damping noise (see Sec.~\ref{sec:adc}), the ancilla-deficient state does not provide a notable advantage over the parallel-entangled strategy, even in the high noise regime.

For the depolarizing channel, the best classical strategy also has QFI
\begin{equation}
J(\mathcal{E}[(\rho^{\otimes{N}})]) = N (1-\eta)^{2}.
\label{eq:depolclassical}
\end{equation}

The QFI ratios of the ancilla-assisted state to \eqref{eq:depolclassical}, to the intermediate strategy and to the optimal 4-qubit state are shown in Fig.~\ref{f:twodepol} (b).

\section{Amplitude damping \label{sec:adc}}

In the presence of amplitude damping noise, we show that the crossover occurs for both the small $N$ regime and the asymptotic limit of large $N$. We also detail the interesting case for $N=2$ where the transition occurs at efficiency $0.5$ exactly. 

The channel is described by the Kraus operators
\begin{equation}
 K_1= \left(
  \begin{array}{cc}
   1 & 0 \\
   0 & \sqrt{1-\eta } \\
  \end{array}
  \right),
K_2= \left(
\begin{array}{cc}
 0 & \sqrt{\eta } \\
 0 & 0 \\
\end{array}
\right).
\end{equation}
\noindent where $\eta$ is the probability that the probe undergoes a transition \mbox{$\ket{1} \rightarrow \ket{0}$}.
For all $\eta$, the optimal ancilla-assisted state takes the form
\mbox{$\alpha(\eta)\ket{00}+ \sqrt{1-\alpha(\eta)^2}\ket{11}$}. Here
 \mbox{$\alpha(\eta)=[\sqrt{1-\eta}/(1+\sqrt{1-\eta})]^{1/2}$} maximizes the QFI $J_\text{adc}$, giving 
\mbox{$J_\text{adc}= 8 (1-\eta)/ \left(\sqrt{1-\eta }+1\right)^2$}.

For all $N$, it was noted in the supplementary material in Ref.~\cite{rafal} that for the parallel-entangled strategy, the upper bound is $N\eta/(1-\eta)$, and the transition occurs at $(1-\eta) \approx 0.36$.

\begin{figure}[h]
\includegraphics[trim = 0 0 0 0, clip, width=0.9\linewidth]{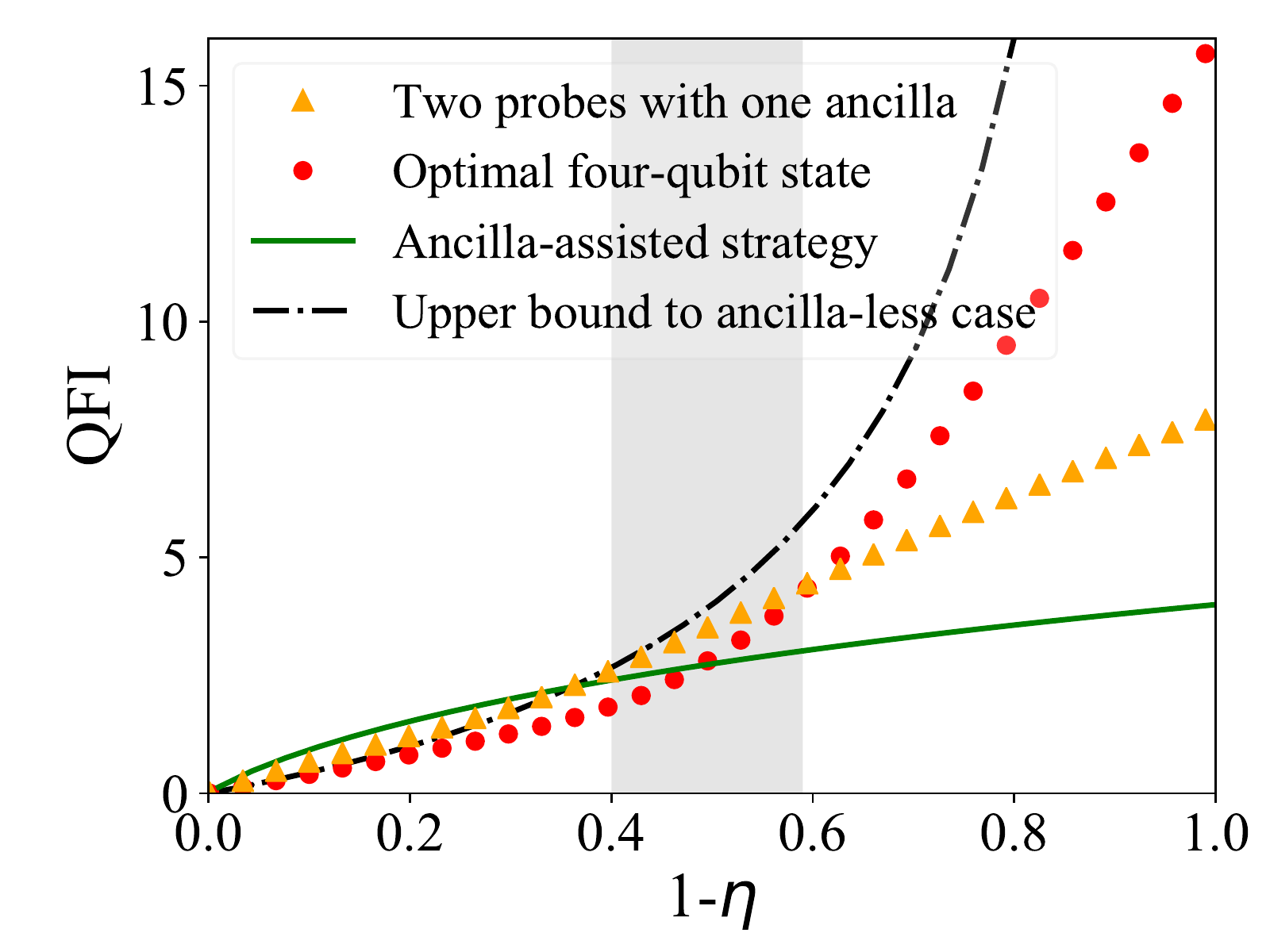}
\includegraphics[trim = 0 0 0 0, clip, width=0.9\linewidth]{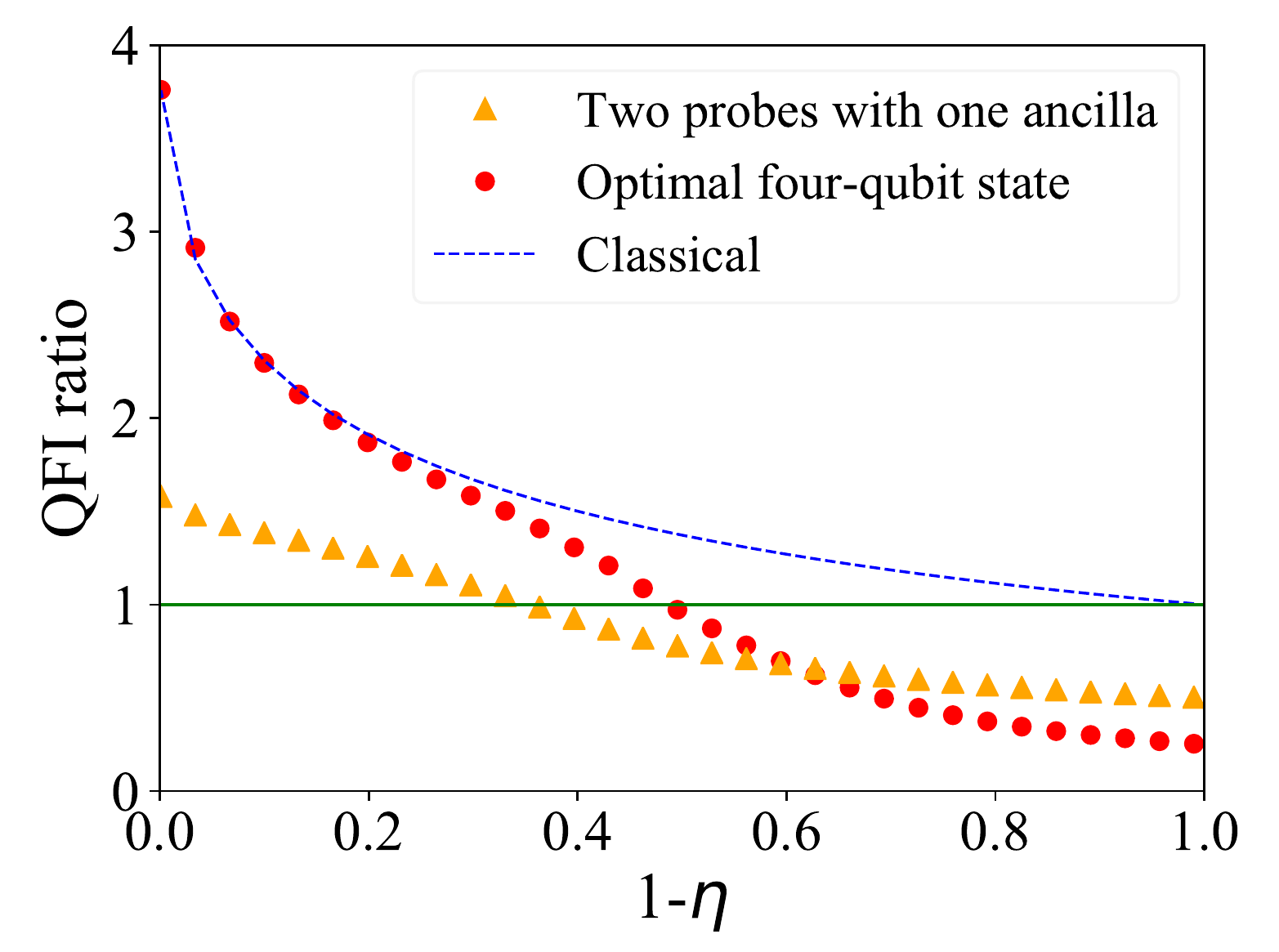}
\caption{\label{f:amp_damp_4qubits} (a) Amplitude damping noise:  QFI for the ancilla-assisted state $\frac{1}{\sqrt{2}}(\ket{00}+\ket{11})$ where one qubit is the probe and the other is an ancilla (green solid line), the optimal parallel-entangled four-qubit state (red circles), the optimal three-qubit state where two are probes and one is an ancilla (orange triangles) and the upper bound to the ancilla-less case (black dashed line) in the presence of amplitude damping noise. The shaded region denotes where the ancilla-deficient state is optimal; to the left of the region, the ancilla-assisted strategy is optimal, and to the right, the parallel-entangled strategy is.
(b) The QFI ratio of the ancilla-assisted strategy to the optimal four-qubit state (red circles), the ancilla-assisted strategy to the intermediate state (yellow triangles), and the ancilla-assisted strategy to the best classical strategy (blue dashed line).}
\end{figure}

As in the Pauli x-y noise, here we also observe two transitions: the ancilla-assisted strategy is optimal in the interval \mbox{$(1-\eta) \in [0,0.4]$}, the intermediate strategy in \mbox{$ [0.4,0.6]$} and the parallel-entangled strategy in 
\mbox{$ [0.6,1.0]$}.
As expected, the intermediate strategy crosses the parallel-entangled strategy at larger efficiencies, but in the noiseless limit, it also out-performs the strategy (b) by a factor of $N/2$.

We also observed that as $N$ is increased from 2 to 6, the transition shifts towards smaller values of $(1-\eta)$, which is consistent with the bound.

For the amplitude damping channel, the optimal classical state is once again $\ket{+}$, with QFI $1-\eta$. The QFI ratios of the ancilla-assisted state to $(1-\eta)$, to the intermediate strategy and to the optimal 4-qubit state are shown in Fig.~\ref{f:amp_damp_4qubits} (b). 

Now we examine the $N=2$ case in more detail, where we see that the cross-over between the parallel-entangled strategy and the ancilla-assisted strategy occurs at exactly $\eta=1/2$. For the parallel-entangled strategy, up to $\eta \approx 0.58$, we observe numerically that the best two-probe state takes the form $\epsilon(\eta)\ket{00}+ \sqrt{1-\epsilon(\eta)^2}\ket{11}$, where
\begin{equation}
\epsilon = \frac{\sqrt{\frac{\eta  (\eta  (2 (\eta -3) \eta +7)-4)-\sqrt{(\eta -1)^4 (2 (\eta -1) \eta +1)}+1}{(\eta -1)^3 \eta }}}{\sqrt{2}}
\end{equation}
\noindent gives the maximum QFI
\begin{equation}
J_\text{pa}= \frac{8 \left(\eta ^2-\sqrt{2 \eta ^2-2 \eta +1}-\eta +1\right)}{(\eta -1)^2}.
\end{equation}

\begin{figure}[hbt]
\includegraphics[trim = 2cm 3cm 3cm 1.7cm, clip, width=0.8\linewidth]{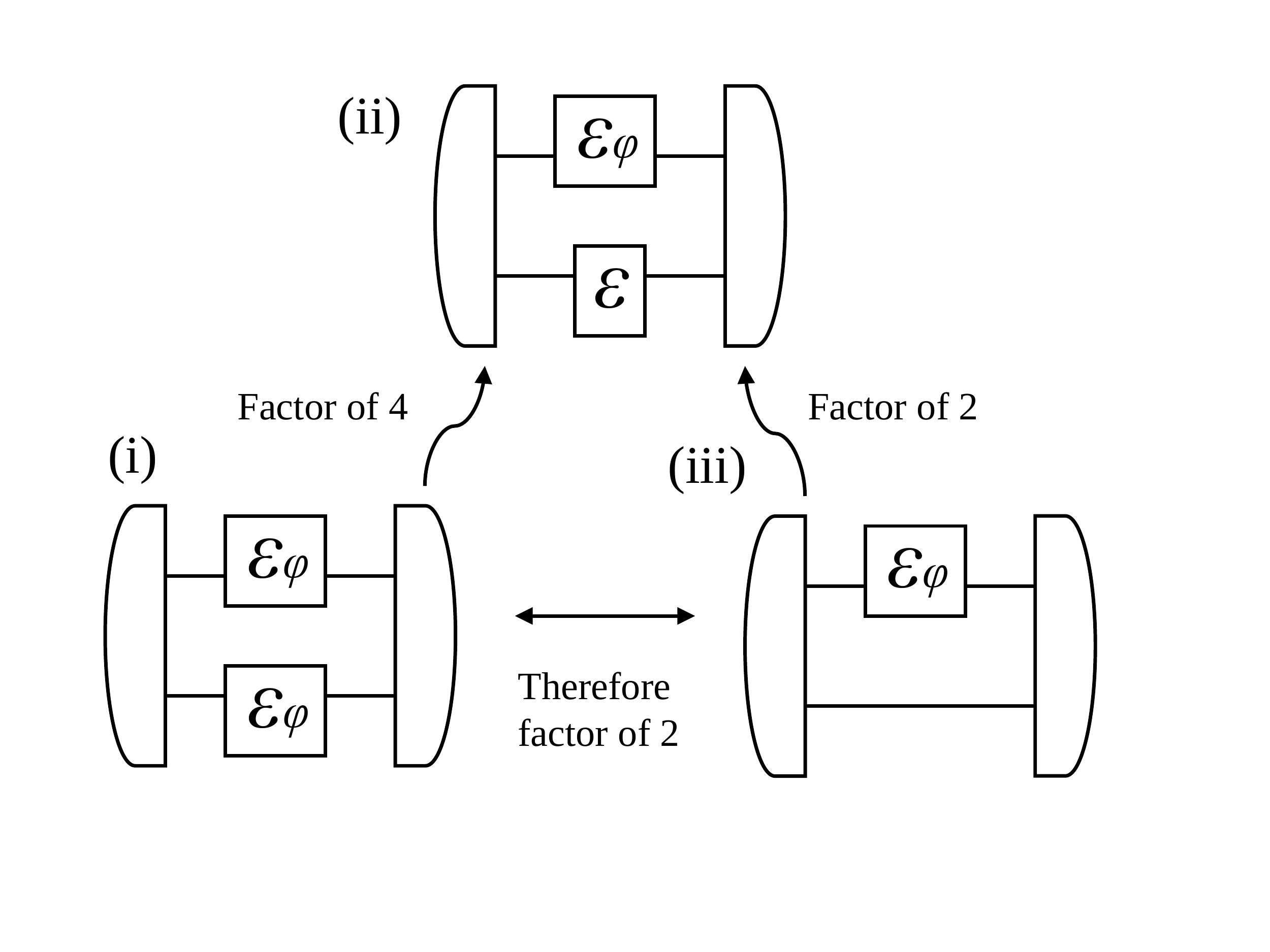}
\caption{\label{f:crossover}  
The amplitude damping channel at $\eta = 0.5$, $\mathcal{E}_\varphi$ refers to applying the noise as well as the phase unitary. $\mathcal{E}$ by itself means that only the noise acts on the probe and no unitary is applied. The schematic shows (i) the parallel-entangled state and (iii) the ancilla-assisted state.}
\end{figure}

At $\eta = 0.5$, $\alpha(\eta)$ coincides with $\epsilon(\eta)$, i.e. the optimal states for the two strategies are the same. This means that using the ancilla-assisted state twice has the same QFI as using the parallel strategy once, and the relationship between the two is shown as a schematic in Fig.~\ref{f:crossover}.
There is a factor of 4 between the QFI of the two-probe state as depicted in (i) and (ii), since the only difference between the two output states is in the off-diagonal component, where the former (latter) picks up a factor of $e^{2 i\varphi}$ ($e^{i \varphi}$).

Now, we compare cases (iii) and (ii) of Fig.~\ref{f:crossover}. The presence of the ancilla acts like a ``filter" for the damped component, it prevents the noise from mixing with the phase sensitive component. Thus, the noise has a similar effect on the probe state as a loss. Acting a lossy channel with transmissivity $(1-\eta)$ reduces its QFI by a factor of $(1-\eta)$. Therefore there is a factor of 2 between (iii) and (ii), hence when (iii) is used twice, it has the same QFI as the two-probe entangled state.

\section{Conclusion \label{sec:con}}

In conclusion, we have shown that in the presence of Pauli x-y, depolarizing and amplitude damping noise, the optimal entanglement-assisted strategy for quantum metrology depends on the strength of the noise parameter. When comparing the parallel-entangled strategy to the ancilla-assisted unentangled strategy, the former is optimal in the low noise regime, since entanglement between probes allows estimation precision to beat the standard quantum limit. In the high noise regime, unentangled probes perform better because they are less noise-sensitive, and in this regime, ancilla-assisted states provide an advantage.

\section{Acknowledgements}
We thank Andrea Smirne, Janek Ko{\l}ody{\'n}ski and Gabriel Durkin for insightful discussions.
% \clearpage

\appendix
\renewcommand\thefigure{\thesection.\arabic{figure}}    
\setcounter{figure}{0} 

% \newrefsection

% 

\section{General phase covariant noise \label{app:a}}
The most general form of phase covariant noise maps have Kraus operators are given by \cite{PhysRevLett.116.120801}:

\begin{eqnarray}
K_1= \sqrt{\frac{1-\eta_{||}+ \kappa}{2}} \left(\begin{array}{cc}
                           0 & 1 \\
                           0 & 0 \\
                          \end{array}
                                       \right) ,        \nonumber \\
K_2= \sqrt{\frac{1-\eta_{||}- \kappa}{2}}  \left(\begin{array}{cc}
                               0 & 0 \\
                               1 & 0 \\
                              \end{array}
                      \right), \nonumber   \\
 K_3= \sqrt{\lambda_+} \left(\begin{array}{cc}
                   \cos(t) & 0 \\
                   0 & \sin(t) \\
                  \end{array}
              \right)     ,              \nonumber \\
  K_4= \sqrt{\lambda_+}    \left(\begin{array}{cc}
                   -\sin(t) & 0 \\
                   0 & \cos(t) \\
                  \end{array}
          \right)         ,                 
\end{eqnarray}
\noindent where 
\begin{eqnarray}
t = \tan^{-1}\left( \frac{2 \eta_\bot}{\kappa+\sqrt{\kappa^2 + 4 \eta_\bot ^2}} \right) \nonumber \\
\lambda \pm = \frac{1+\eta_{||} \pm \sqrt{\kappa^2 + 4 \eta_{\bot} ^2}  }{2} .
\end{eqnarray}

\noindent For the map to be complete positive and trace preserving, these conditions are required:
\begin{eqnarray}
\eta_{||} + \kappa \leq 1, \hspace{5mm} 1+\eta_{||} \geq \sqrt{4 \eta_\bot ^2 + \kappa^2}
\label{eq:cptpcond}
\end{eqnarray}

In the Bloch sphere representation, $|\kappa| \neq 0 $ corresponds to a displacement of qubit in the $z$ direction and 
\mbox{$-1\leq \kappa \leq 1$}; $\eta_{\bot}$ represents its length in the $x-y$ plane, with $0 < \eta_{\bot} < 1$, and $\eta_{||}$ is its length in the $z$ direction, with $0 \leq \eta_{||} \leq 1 $.
Pure dephasing corresponds to $\eta_{||}=1,\kappa=0$ and $0 \leq \eta_\bot \leq 1$.  
Relaxation (excitation) corresponds to \mbox{$0 \leq \kappa \leq 1$} (\mbox{$-1 \leq \kappa \leq 0$}). Isotropic depolarization is given by $0 \leq \eta_{||}= \eta_\bot \leq 1$.

Since we know that ancillae do not provide an advantage in the small $N$ limit, to obtain the biggest difference in QFI between the two strategies as a function of $\kappa$ and $\eta_{||}$, $\eta_\bot$ is chosen to be $\sqrt{1+\eta_{||}^2 - \kappa^2}/2$, i.e., the largest value possible given by the conditions in Eq.~\eqref{eq:cptpcond}.

\begin{figure}[tbh]\vspace{-0.5cm}
\includegraphics[trim = 0cm 0.3cm 0.2cm 1cm, clip, width=0.9\linewidth]{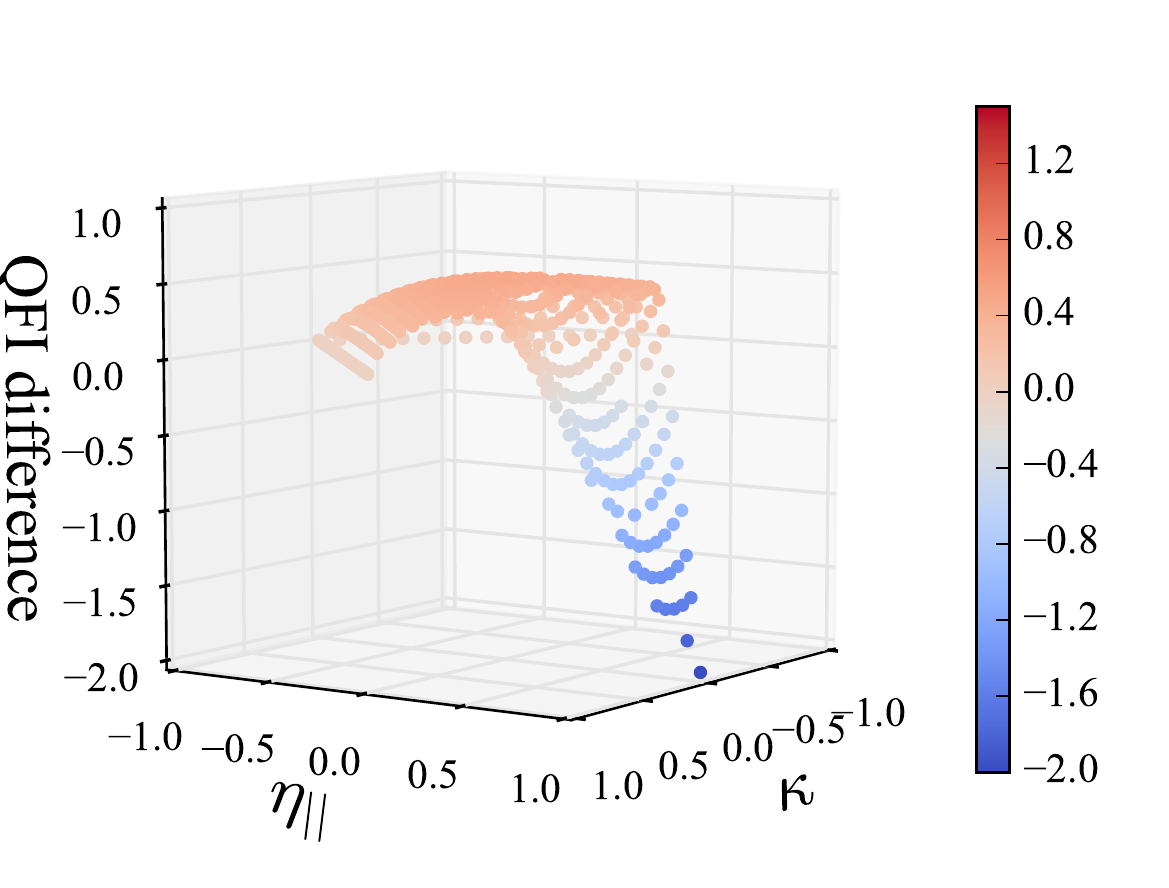}
\includegraphics[trim = 1.1cm 0cm 0.5cm 1cm, clip, width=0.9\linewidth]{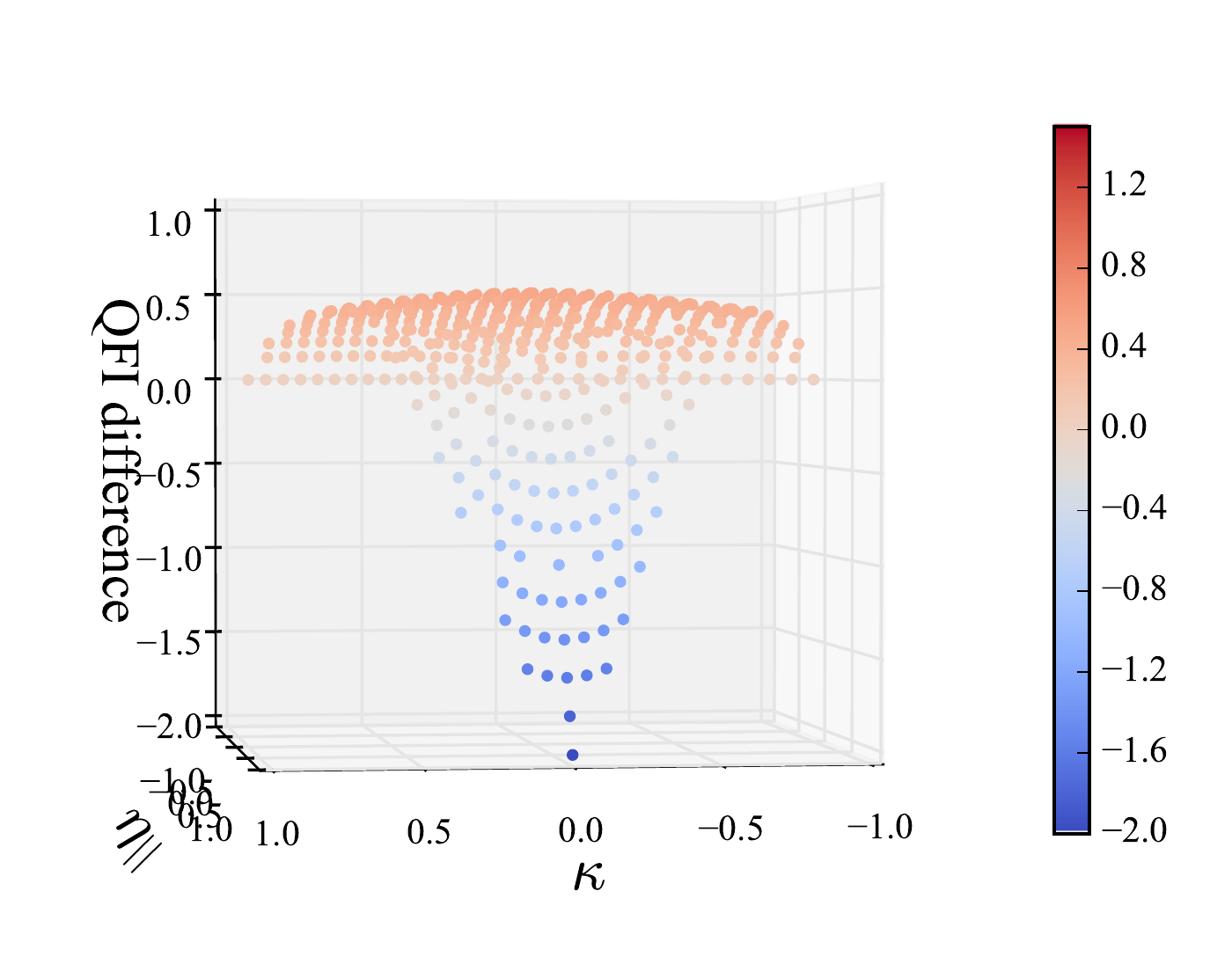}
\caption{\label{f:xy}  The phase-covariant noise channel: for $N=2$, the QFI difference between the optimal ancilla-assisted state and the optimal parallel-entangled state. We show the plot from different perspectives for clarity.}
\end{figure}

For $N=2$, the difference in QFI between the optimal parallel-entangled state and the ancilla-assisted state is plotted in Fig.~\ref{f:xy}. Values above zero means that the ancilla-assisted strategy performs better.
 We see that there is a large parameter space for which the data points are positive, i.e. the ancilla-assisted state is advantageous. As expected, the difference is the most negative (ie the parallel-entangled strategy performs the best) when $\kappa = 0, \eta_{||} =1$, corresponding to the noiseless regime.

\section{Noise parameter estimation \label{app:b}}

The crossover between the strategies occurs because in the noiseless limit, the quantum Fisher information scales as $N^2$, which occurs only for unitary transformations \cite{fujiwara2008fibre}. For a of class channels which authors in Ref.~\cite{PhysRevLett.118.100502} termed ``teleportation stretchable", i.e., the channels which commute with teleportation, they show that the estimation cannot exceed the standard quantum limit, and the highest quantum Fisher information is given by the Choi matrix of the channel. This coincides with the ancilla-assisted scheme in Fig.~\ref{f:scheme} (b).

Here, for estimating the depolarizing parameter, we investigate the performance the different quantum metrology strategies in Ref.~\cite{rafal}. We also numerically confirm the result in Ref.~\cite{PhysRevLett.118.100502}.
Firstly, we rewrite Eq.~\eqref{eq:pauli} as
\begin{equation}
\mathcal{E} (\rho) \rightarrow (1-t) \rho + t \frac{\openone}{d}.
\end{equation}

\noindent  In the cases where analytical solutions can be obtained, we use Fisher information to characterize the achievable precision:

\begin{equation}
J(\rho_t) = \displaystyle \sum_i \lambda(i|t) \left(\frac{\partial \log[\lambda(i|t) ]}{\partial t} \right)^2
\end{equation}
where $\lambda(i|t)$ correspond to the eigenvalues of the density matrix $\rho_t$ at the output of the channel. 

We start with the sequential strategy, where an unentangled probe is cycled the channel $N$ time sequentially. We parametrize the input state as $\epsilon\ket{0}+\sqrt{1-\epsilon^2}\ket{0}$. The eigenvalues of the output density matrix are  $1-\frac{t}{2}$ and $ \frac{t}{2}$, and the eigenvectors are independent of $t$. In this case, the Fisher information is $J(\mathcal{E}(\rho_\text{seq}))  =  1/(2 t - t^2)$. Applying the channel twice gives Fisher information 
\begin{equation}
J[\mathcal{E}^2(\rho_\text{seq})] = \frac{4 (p-1)^2}{(2-p) p \left(p^2-2 p+2\right)}
\end{equation}

\noindent and three times 

\begin{equation}
J[\mathcal{E}^3(\rho_\text{seq})] =-\frac{9 (p-1)^4}{p \left(p^2-3 p+3\right) \left(p^3-3 p^2+3 p-2\right)}
\end{equation}

\begin{figure}[h!]
\includegraphics[trim = 0 0 0 0, clip, width=0.9\linewidth]{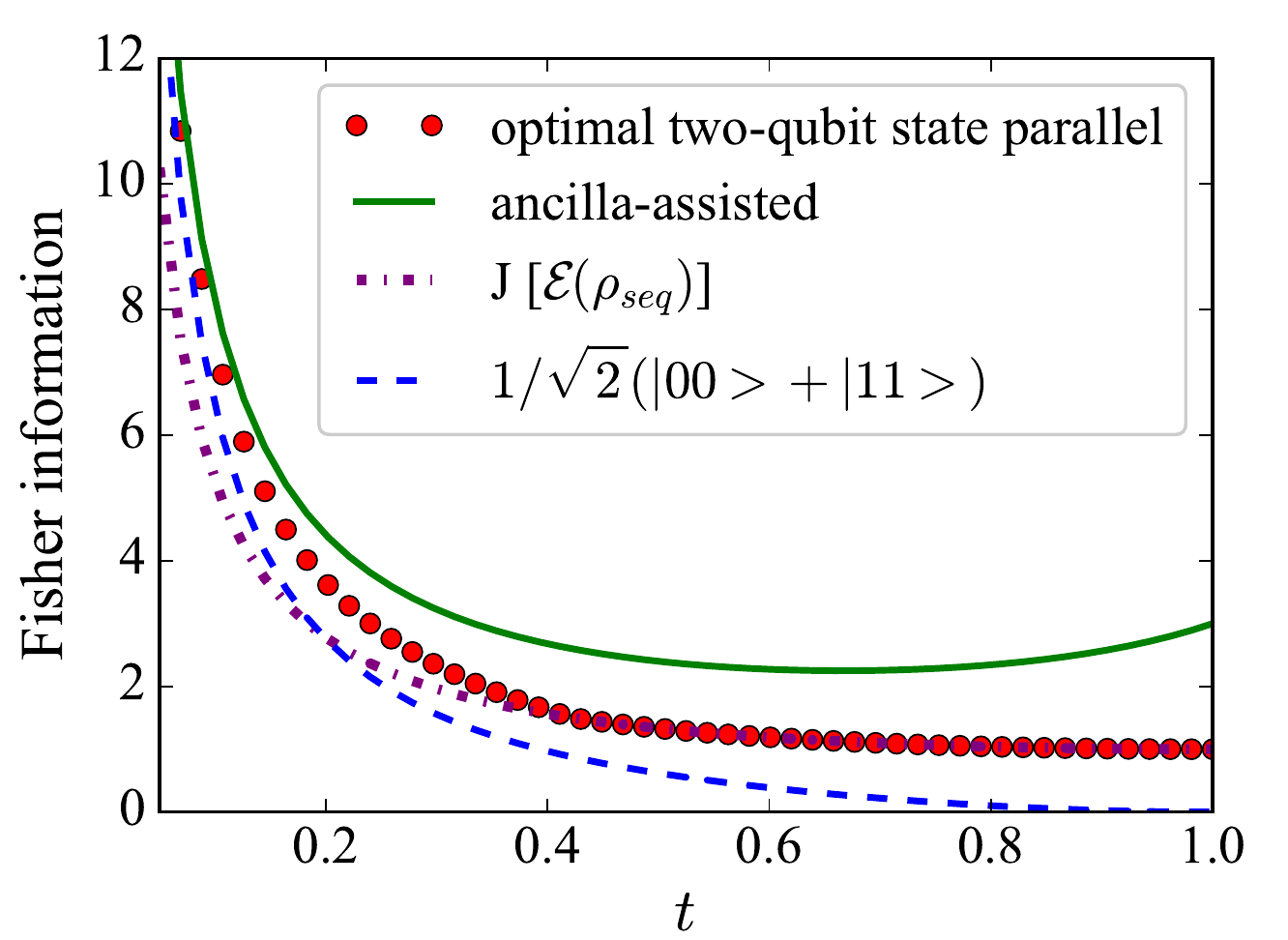}
\caption{\label{f:two_qubits_depol} QFI for two uses of the channel, estimating the depolarizing probability. The results shown are: the sequential strategy $J[\mathcal{E}(\rho_\text{seq})]$ (purple dotted dashed line), the maximally entangled state where both are probes (dashed blue line), the optimal two-probe state (red circles) and the ancilla-assisted state, where one probe is maximally to a noiseless ancilla (solid green line).}
\end{figure}

We observe that $ J[\mathcal{E}(\rho_\text{seq})] > J[\mathcal{E}^2(\rho_\text{seq})]/2 > J[\mathcal{E}^3(\rho_\text{seq})]/3$. Therefore using unentangled probes, if the channel can be sampled $N$ times, the best precision is achieved by using $N$ probes to go through the channel once. We show $J[\mathcal{E}(\rho_\text{seq})]$ in Fig.~\ref{f:two_qubits_depol} as a purple dotted-dashed line.

Now we turn to the parallel-entangled strategy, where $N$ probes go through the channel in parallel. We show that using entanglement can improve the precision of the estimation. 
We numerically search for the optimal two-probe state. Since the transformation is non-unitary, the definition for QFI in Eq.~\eqref{eq:QFI} cannot be used since the generator of the operation of the channel is not trivial to define.  In this case, we use an alternative definition:
\begin{equation}
J(t) = \frac{8(1-F[\rho(t),\rho(t+\delta t )])}{d t^2}
\label{eq:alt}
\end{equation}
\noindent where F$(\rho,\sigma) = $Tr$\sqrt{\sqrt{\sigma} \rho \sqrt{\sigma}}$ is the Bure's quantum fidelity between $\rho$ and $\sigma$ \cite{hubner1992explicit,uhlmann1976transition,jozsa1994fidelity}. 

Since the QFI of the state in Eq.~\eqref{eq:alt} is smooth in $t$, we choose a reasonably small $\delta t \rightarrow 0$, and find the optimal two-qubit probe state by numerically optimizing 
\begin{equation}
\frac{8(1-F[\rho(t- \frac{\delta t}{2}),\rho(p+ \frac{\delta t}{2})])}{\delta t^2}.
\end{equation}
\noindent The solution is shown by the red circles.

For comparison, the maximally entangled state (both are probes) has QFI given by
\begin{eqnarray}
J_\text{NOON} = \frac{3}{4} (t-1)^2 \left(2 t-t^2\right) \left(\frac{3}{3 t^2-6 t+4}+\frac{4}{(t-2)^2 t^2}\right) \nonumber \\
\end{eqnarray}
\noindent and we plot $J_\text{NOON}/2$ as a dashed blue line in Fig.~\ref{f:two_qubits_depol}.

The ancilla-assisted state $1/\sqrt{2}(\ket{00}+\ket{11})$, where one is the probe and the other is a noiseless ancilla has QFI $J_\text{anc} = 3/(4t - 3 t^2 )$, which is plotted as a solid green line.

As evident from the plot, the maximally entangled state performs better than the sequential strategy when $t \leq 0.2$, and the optimal two-probe parallel-entangled state starts to out-perform the unentangled probe around  $t \leq 0.4$. Clearly, the parallel-entangled strategy must perform better than the unentangled strategy, because the separable states are a subset of the parallel-entangled ones. The ancilla-assisted scheme, where one probe is maximally entangled to an ancilla has the highest quantum Fisher information.

We also performed a search over all possible four-qubit states where two are probes and two are ancillae. The optimization shows that the optimal state is still $1/\sqrt{2}(\ket{00}+\ket{11})$ used twice - which confirms that for these channels, indeed one cannot perform better than using the ancilla-assisted state.

To conclude, the parallel-entangled strategy performs better than the sequential strategy (by definition, since the latter is a special case of the former), and the ultimate precision is achieved by the ancilla-assisted strategy, where the probe is maximally entangled with an ancilla.


\begin{thebibliography}{24}%
\makeatletter
\providecommand \@ifxundefined [1]{%
 \@ifx{#1\undefined}
}%
\providecommand \@ifnum [1]{%
 \ifnum #1\expandafter \@firstoftwo
 \else \expandafter \@secondoftwo
 \fi
}%
\providecommand \@ifx [1]{%
 \ifx #1\expandafter \@firstoftwo
 \else \expandafter \@secondoftwo
 \fi
}%
\providecommand \natexlab [1]{#1}%
\providecommand \enquote  [1]{``#1''}%
\providecommand \bibnamefont  [1]{#1}%
\providecommand \bibfnamefont [1]{#1}%
\providecommand \citenamefont [1]{#1}%
\providecommand \href@noop [0]{\@secondoftwo}%
\providecommand \href [0]{\begingroup \@sanitize@url \@href}%
\providecommand \@href[1]{\@@startlink{#1}\@@href}%
\providecommand \@@href[1]{\endgroup#1\@@endlink}%
\providecommand \@sanitize@url [0]{\catcode `\\12\catcode `\$12\catcode
  `\&12\catcode `\#12\catcode `\^12\catcode `\_12\catcode `\%12\relax}%
\providecommand \@@startlink[1]{}%
\providecommand \@@endlink[0]{}%
\providecommand \url  [0]{\begingroup\@sanitize@url \@url }%
\providecommand \@url [1]{\endgroup\@href {#1}{\urlprefix }}%
\providecommand \urlprefix  [0]{URL }%
\providecommand \Eprint [0]{\href }%
\providecommand \doibase [0]{http://dx.doi.org/}%
\providecommand \selectlanguage [0]{\@gobble}%
\providecommand \bibinfo  [0]{\@secondoftwo}%
\providecommand \bibfield  [0]{\@secondoftwo}%
\providecommand \translation [1]{[#1]}%
\providecommand \BibitemOpen [0]{}%
\providecommand \bibitemStop [0]{}%
\providecommand \bibitemNoStop [0]{.\EOS\space}%
\providecommand \EOS [0]{\spacefactor3000\relax}%
\providecommand \BibitemShut  [1]{\csname bibitem#1\endcsname}%
\let\auto@bib@innerbib\@empty
%</preamble>
\bibitem [{\citenamefont {Giovannetti}\ \emph {et~al.}(2004)\citenamefont
  {Giovannetti}, \citenamefont {Lloyd},\ and\ \citenamefont
  {Maccone}}]{giovannetti2004quantum}%
  \BibitemOpen
  \bibfield  {author} {\bibinfo {author} {\bibfnamefont {V.}~\bibnamefont
  {Giovannetti}}, \bibinfo {author} {\bibfnamefont {S.}~\bibnamefont {Lloyd}},
  \ and\ \bibinfo {author} {\bibfnamefont {L.}~\bibnamefont {Maccone}},\
  }\href@noop {} {\bibfield  {journal} {\bibinfo  {journal} {Science}\ }\textbf
  {\bibinfo {volume} {306}},\ \bibinfo {pages} {1330} (\bibinfo {year}
  {2004})}\BibitemShut {NoStop}%
\bibitem [{\citenamefont {Giovannetti}\ \emph {et~al.}(2006)\citenamefont
  {Giovannetti}, \citenamefont {Lloyd},\ and\ \citenamefont
  {Maccone}}]{PhysRevLett.96.010401}%
  \BibitemOpen
  \bibfield  {author} {\bibinfo {author} {\bibfnamefont {V.}~\bibnamefont
  {Giovannetti}}, \bibinfo {author} {\bibfnamefont {S.}~\bibnamefont {Lloyd}},
  \ and\ \bibinfo {author} {\bibfnamefont {L.}~\bibnamefont {Maccone}},\ }\href
  {\doibase 10.1103/PhysRevLett.96.010401} {\bibfield  {journal} {\bibinfo
  {journal} {Phys. Rev. Lett.}\ }\textbf {\bibinfo {volume} {96}},\ \bibinfo
  {pages} {010401} (\bibinfo {year} {2006})}\BibitemShut {NoStop}%
\bibitem [{\citenamefont {van Dam}\ \emph {et~al.}(2007)\citenamefont {van
  Dam}, \citenamefont {D'Ariano}, \citenamefont {Ekert}, \citenamefont
  {Macchiavello},\ and\ \citenamefont {Mosca}}]{PhysRevLett.98.090501}%
  \BibitemOpen
  \bibfield  {author} {\bibinfo {author} {\bibfnamefont {W.}~\bibnamefont {van
  Dam}}, \bibinfo {author} {\bibfnamefont {G.~M.}\ \bibnamefont {D'Ariano}},
  \bibinfo {author} {\bibfnamefont {A.}~\bibnamefont {Ekert}}, \bibinfo
  {author} {\bibfnamefont {C.}~\bibnamefont {Macchiavello}}, \ and\ \bibinfo
  {author} {\bibfnamefont {M.}~\bibnamefont {Mosca}},\ }\href {\doibase
  10.1103/PhysRevLett.98.090501} {\bibfield  {journal} {\bibinfo  {journal}
  {Phys. Rev. Lett.}\ }\textbf {\bibinfo {volume} {98}},\ \bibinfo {pages}
  {090501} (\bibinfo {year} {2007})}\BibitemShut {NoStop}%
\bibitem [{\citenamefont {Demkowicz-Dobrza\ifmmode~\acute{n}\else
  \'{n}\fi{}ski}\ and\ \citenamefont {Maccone}(2014)}]{rafal}%
  \BibitemOpen
  \bibfield  {author} {\bibinfo {author} {\bibfnamefont {R.}~\bibnamefont
  {Demkowicz-Dobrza\ifmmode~\acute{n}\else \'{n}\fi{}ski}}\ and\ \bibinfo
  {author} {\bibfnamefont {L.}~\bibnamefont {Maccone}},\ }\href {\doibase
  10.1103/PhysRevLett.113.250801} {\bibfield  {journal} {\bibinfo  {journal}
  {Phys. Rev. Lett.}\ }\textbf {\bibinfo {volume} {113}},\ \bibinfo {pages}
  {250801} (\bibinfo {year} {2014})}\BibitemShut {NoStop}%
\bibitem [{\citenamefont {Demkowicz-Dobrza{\'n}ski}\ \emph
  {et~al.}(2012)\citenamefont {Demkowicz-Dobrza{\'n}ski}, \citenamefont
  {Ko{\l}ody{\'n}ski},\ and\ \citenamefont {Gu{\c{t}}{\u{a}}}}]{guta}%
  \BibitemOpen
  \bibfield  {author} {\bibinfo {author} {\bibfnamefont {R.}~\bibnamefont
  {Demkowicz-Dobrza{\'n}ski}}, \bibinfo {author} {\bibfnamefont
  {J.}~\bibnamefont {Ko{\l}ody{\'n}ski}}, \ and\ \bibinfo {author}
  {\bibfnamefont {M.}~\bibnamefont {Gu{\c{t}}{\u{a}}}},\ }\href@noop {}
  {\bibfield  {journal} {\bibinfo  {journal} {Nat. Comm.}\ }\textbf {\bibinfo
  {volume} {3}},\ \bibinfo {pages} {1063} (\bibinfo {year} {2012})}\BibitemShut
  {NoStop}%
\bibitem [{\citenamefont {Dowling}(2008)}]{Dowling2008}%
  \BibitemOpen
  \bibfield  {author} {\bibinfo {author} {\bibfnamefont {J.~P.}\ \bibnamefont
  {Dowling}},\ }\href@noop {} {\bibfield  {journal} {\bibinfo  {journal}
  {Contemp. Phys.}\ }\textbf {\bibinfo {volume} {49}},\ \bibinfo {pages} {125}
  (\bibinfo {year} {2008})}\BibitemShut {NoStop}%
\bibitem [{\citenamefont {Dorner}\ \emph {et~al.}(2009)\citenamefont {Dorner},
  \citenamefont {Demkowicz-Dobrzanski}, \citenamefont {Smith}, \citenamefont
  {Lundeen}, \citenamefont {Wasilewski}, \citenamefont {Banaszek},\ and\
  \citenamefont {Walmsley}}]{prl102040403}%
  \BibitemOpen
  \bibfield  {author} {\bibinfo {author} {\bibfnamefont {U.}~\bibnamefont
  {Dorner}}, \bibinfo {author} {\bibfnamefont {R.}~\bibnamefont
  {Demkowicz-Dobrzanski}}, \bibinfo {author} {\bibfnamefont {B.~J.}\
  \bibnamefont {Smith}}, \bibinfo {author} {\bibfnamefont {J.~S.}\ \bibnamefont
  {Lundeen}}, \bibinfo {author} {\bibfnamefont {W.}~\bibnamefont {Wasilewski}},
  \bibinfo {author} {\bibfnamefont {K.}~\bibnamefont {Banaszek}}, \ and\
  \bibinfo {author} {\bibfnamefont {I.~A.}\ \bibnamefont {Walmsley}},\ }\href
  {\doibase 10.1103/PhysRevLett.102.040403} {\bibfield  {journal} {\bibinfo
  {journal} {Phys. Rev. Lett.}\ }\textbf {\bibinfo {volume} {102}},\ \bibinfo
  {pages} {040403} (\bibinfo {year} {2009})}\BibitemShut {NoStop}%
\bibitem [{\citenamefont {Dinani}\ and\ \citenamefont
  {Berry}(2014{\natexlab{a}})}]{dinani}%
  \BibitemOpen
  \bibfield  {author} {\bibinfo {author} {\bibfnamefont {H.~T.}\ \bibnamefont
  {Dinani}}\ and\ \bibinfo {author} {\bibfnamefont {D.~W.}\ \bibnamefont
  {Berry}},\ }\href {\doibase 10.1103/PhysRevA.90.023856} {\bibfield  {journal}
  {\bibinfo  {journal} {Phys. Rev. A}\ }\textbf {\bibinfo {volume} {90}},\
  \bibinfo {pages} {023856} (\bibinfo {year} {2014}{\natexlab{a}})}\BibitemShut
  {NoStop}%
\bibitem [{\citenamefont {Ko{\l}ody{\'n}ski}\ and\ \citenamefont
  {Demkowicz-Dobrza{\'n}ski}(2013)}]{kolodynski2013efficient}%
  \BibitemOpen
  \bibfield  {author} {\bibinfo {author} {\bibfnamefont {J.}~\bibnamefont
  {Ko{\l}ody{\'n}ski}}\ and\ \bibinfo {author} {\bibfnamefont {R.}~\bibnamefont
  {Demkowicz-Dobrza{\'n}ski}},\ }\href@noop {} {\bibfield  {journal} {\bibinfo
  {journal} {N. J. Phys.}\ }\textbf {\bibinfo {volume} {15}},\ \bibinfo {pages}
  {073043} (\bibinfo {year} {2013})}\BibitemShut {NoStop}%
\bibitem [{\citenamefont {Huelga}\ \emph {et~al.}(1997)\citenamefont {Huelga},
  \citenamefont {Macchiavello}, \citenamefont {Pellizzari}, \citenamefont
  {Ekert}, \citenamefont {Plenio},\ and\ \citenamefont {Cirac}}]{chiara}%
  \BibitemOpen
  \bibfield  {author} {\bibinfo {author} {\bibfnamefont {S.~F.}\ \bibnamefont
  {Huelga}}, \bibinfo {author} {\bibfnamefont {C.}~\bibnamefont
  {Macchiavello}}, \bibinfo {author} {\bibfnamefont {T.}~\bibnamefont
  {Pellizzari}}, \bibinfo {author} {\bibfnamefont {A.~K.}\ \bibnamefont
  {Ekert}}, \bibinfo {author} {\bibfnamefont {M.~B.}\ \bibnamefont {Plenio}}, \
  and\ \bibinfo {author} {\bibfnamefont {J.~I.}\ \bibnamefont {Cirac}},\ }\href
  {\doibase 10.1103/PhysRevLett.79.3865} {\bibfield  {journal} {\bibinfo
  {journal} {Phys. Rev. Lett.}\ }\textbf {\bibinfo {volume} {79}},\ \bibinfo
  {pages} {3865} (\bibinfo {year} {1997})}\BibitemShut {NoStop}%
\bibitem [{\citenamefont {Escher}\ \emph {et~al.}(2011)\citenamefont {Escher},
  \citenamefont {de~Matos~Filho},\ and\ \citenamefont
  {Davidovich}}]{davidovich}%
  \BibitemOpen
  \bibfield  {author} {\bibinfo {author} {\bibfnamefont {B.}~\bibnamefont
  {Escher}}, \bibinfo {author} {\bibfnamefont {R.}~\bibnamefont
  {de~Matos~Filho}}, \ and\ \bibinfo {author} {\bibfnamefont {L.}~\bibnamefont
  {Davidovich}},\ }\href@noop {} {\bibfield  {journal} {\bibinfo  {journal}
  {Nat. Phys.}\ }\textbf {\bibinfo {volume} {7}},\ \bibinfo {pages} {406}
  (\bibinfo {year} {2011})}\BibitemShut {NoStop}%
\bibitem [{\citenamefont {Huang}\ \emph {et~al.}(2016)\citenamefont {Huang},
  \citenamefont {Macchiavello},\ and\ \citenamefont
  {Maccone}}]{PhysRevA.94.012101}%
  \BibitemOpen
  \bibfield  {author} {\bibinfo {author} {\bibfnamefont {Z.}~\bibnamefont
  {Huang}}, \bibinfo {author} {\bibfnamefont {C.}~\bibnamefont {Macchiavello}},
  \ and\ \bibinfo {author} {\bibfnamefont {L.}~\bibnamefont {Maccone}},\ }\href
  {\doibase 10.1103/PhysRevA.94.012101} {\bibfield  {journal} {\bibinfo
  {journal} {Phys. Rev. A}\ }\textbf {\bibinfo {volume} {94}},\ \bibinfo
  {pages} {012101} (\bibinfo {year} {2016})}\BibitemShut {NoStop}%
\bibitem [{\citenamefont {Dinani}\ and\ \citenamefont
  {Berry}(2014{\natexlab{b}})}]{PhysRevA.90.023856}%
  \BibitemOpen
  \bibfield  {author} {\bibinfo {author} {\bibfnamefont {H.~T.}\ \bibnamefont
  {Dinani}}\ and\ \bibinfo {author} {\bibfnamefont {D.~W.}\ \bibnamefont
  {Berry}},\ }\href {\doibase 10.1103/PhysRevA.90.023856} {\bibfield  {journal}
  {\bibinfo  {journal} {Phys. Rev. A}\ }\textbf {\bibinfo {volume} {90}},\
  \bibinfo {pages} {023856} (\bibinfo {year} {2014}{\natexlab{b}})}\BibitemShut
  {NoStop}%
\bibitem [{\citenamefont {Smirne}\ \emph {et~al.}(2016)\citenamefont {Smirne},
  \citenamefont {Ko\l{}ody\ifmmode~\acute{n}\else \'{n}\fi{}ski}, \citenamefont
  {Huelga},\ and\ \citenamefont {Demkowicz-Dobrza\ifmmode~\acute{n}\else
  \'{n}\fi{}ski}}]{PhysRevLett.116.120801}%
  \BibitemOpen
  \bibfield  {author} {\bibinfo {author} {\bibfnamefont {A.}~\bibnamefont
  {Smirne}}, \bibinfo {author} {\bibfnamefont {J.}~\bibnamefont
  {Ko\l{}ody\ifmmode~\acute{n}\else \'{n}\fi{}ski}}, \bibinfo {author}
  {\bibfnamefont {S.~F.}\ \bibnamefont {Huelga}}, \ and\ \bibinfo {author}
  {\bibfnamefont {R.}~\bibnamefont {Demkowicz-Dobrza\ifmmode~\acute{n}\else
  \'{n}\fi{}ski}},\ }\href {\doibase 10.1103/PhysRevLett.116.120801} {\bibfield
   {journal} {\bibinfo  {journal} {Phys. Rev. Lett.}\ }\textbf {\bibinfo
  {volume} {116}},\ \bibinfo {pages} {120801} (\bibinfo {year}
  {2016})}\BibitemShut {NoStop}%
\bibitem [{\citenamefont {Holevo}(2011)}]{holevo}%
  \BibitemOpen
  \bibfield  {author} {\bibinfo {author} {\bibfnamefont {A.~S.}\ \bibnamefont
  {Holevo}},\ }\href@noop {} {\emph {\bibinfo {title} {Probabilistic and
  statistical aspects of quantum theory}}}\ (\bibinfo  {publisher} {Springer
  Science \& Business Media},\ \bibinfo {year} {2011})\BibitemShut {NoStop}%
\bibitem [{\citenamefont {Helstrom}(1976)}]{helstrom}%
  \BibitemOpen
  \bibfield  {author} {\bibinfo {author} {\bibfnamefont {C.~W.}\ \bibnamefont
  {Helstrom}},\ }\href@noop {} {\emph {\bibinfo {title} {Quantum detection and
  estimation theory}}}\ (\bibinfo  {publisher} {Academic press},\ \bibinfo
  {year} {1976})\BibitemShut {NoStop}%
\bibitem [{\citenamefont {Braunstein}\ and\ \citenamefont
  {Caves}(1994)}]{caves}%
  \BibitemOpen
  \bibfield  {author} {\bibinfo {author} {\bibfnamefont {S.~L.}\ \bibnamefont
  {Braunstein}}\ and\ \bibinfo {author} {\bibfnamefont {C.~M.}\ \bibnamefont
  {Caves}},\ }\href {\doibase 10.1103/PhysRevLett.72.3439} {\bibfield
  {journal} {\bibinfo  {journal} {Phys. Rev. Lett.}\ }\textbf {\bibinfo
  {volume} {72}},\ \bibinfo {pages} {3439} (\bibinfo {year}
  {1994})}\BibitemShut {NoStop}%
\bibitem [{\citenamefont {Afnan}\ \emph {et~al.}(1996)\citenamefont {Afnan},
  \citenamefont {Banerjee}, \citenamefont {Braunstein}, \citenamefont {Brevik},
  \citenamefont {Caves}, \citenamefont {Chakraborty}, \citenamefont
  {Fischbach}, \citenamefont {Lindblom}, \citenamefont {Milburn}, \citenamefont
  {Odintsov} \emph {et~al.}}]{caves1}%
  \BibitemOpen
  \bibfield  {author} {\bibinfo {author} {\bibfnamefont {I.}~\bibnamefont
  {Afnan}}, \bibinfo {author} {\bibfnamefont {R.}~\bibnamefont {Banerjee}},
  \bibinfo {author} {\bibfnamefont {S.~L.}\ \bibnamefont {Braunstein}},
  \bibinfo {author} {\bibfnamefont {I.}~\bibnamefont {Brevik}}, \bibinfo
  {author} {\bibfnamefont {C.~M.}\ \bibnamefont {Caves}}, \bibinfo {author}
  {\bibfnamefont {B.}~\bibnamefont {Chakraborty}}, \bibinfo {author}
  {\bibfnamefont {E.}~\bibnamefont {Fischbach}}, \bibinfo {author}
  {\bibfnamefont {L.}~\bibnamefont {Lindblom}}, \bibinfo {author}
  {\bibfnamefont {G.}~\bibnamefont {Milburn}}, \bibinfo {author} {\bibfnamefont
  {S.}~\bibnamefont {Odintsov}},  \emph {et~al.},\ }\href@noop {} {\bibfield
  {journal} {\bibinfo  {journal} {Ann. Phys.}\ }\textbf {\bibinfo {volume}
  {247}},\ \bibinfo {pages} {447} (\bibinfo {year} {1996})}\BibitemShut
  {NoStop}%
\bibitem [{\citenamefont {Knysh}\ \emph {et~al.}(2014)\citenamefont {Knysh},
  \citenamefont {Chen},\ and\ \citenamefont {Durkin}}]{1402.0495}%
  \BibitemOpen
  \bibfield  {author} {\bibinfo {author} {\bibfnamefont {S.~I.}\ \bibnamefont
  {Knysh}}, \bibinfo {author} {\bibfnamefont {E.~H.}\ \bibnamefont {Chen}}, \
  and\ \bibinfo {author} {\bibfnamefont {G.~A.}\ \bibnamefont {Durkin}},\
  }\href@noop {} {\bibfield  {journal} {\bibinfo  {journal} {arXiv preprint
  arXiv:1402.0495}\ } (\bibinfo {year} {2014})}\BibitemShut {NoStop}%
\bibitem [{\citenamefont {Fujiwara}\ and\ \citenamefont
  {Imai}(2008)}]{fujiwara2008fibre}%
  \BibitemOpen
  \bibfield  {author} {\bibinfo {author} {\bibfnamefont {A.}~\bibnamefont
  {Fujiwara}}\ and\ \bibinfo {author} {\bibfnamefont {H.}~\bibnamefont
  {Imai}},\ }\href@noop {} {\bibfield  {journal} {\bibinfo  {journal} {J. Phys.
  A}\ }\textbf {\bibinfo {volume} {41}},\ \bibinfo {pages} {255304} (\bibinfo
  {year} {2008})}\BibitemShut {NoStop}%
\bibitem [{\citenamefont {Pirandola}\ and\ \citenamefont
  {Lupo}(2017)}]{PhysRevLett.118.100502}%
  \BibitemOpen
  \bibfield  {author} {\bibinfo {author} {\bibfnamefont {S.}~\bibnamefont
  {Pirandola}}\ and\ \bibinfo {author} {\bibfnamefont {C.}~\bibnamefont
  {Lupo}},\ }\href {\doibase 10.1103/PhysRevLett.118.100502} {\bibfield
  {journal} {\bibinfo  {journal} {Phys. Rev. Lett.}\ }\textbf {\bibinfo
  {volume} {118}},\ \bibinfo {pages} {100502} (\bibinfo {year}
  {2017})}\BibitemShut {NoStop}%
\bibitem [{\citenamefont {H{\"u}bner}(1992)}]{hubner1992explicit}%
  \BibitemOpen
  \bibfield  {author} {\bibinfo {author} {\bibfnamefont {M.}~\bibnamefont
  {H{\"u}bner}},\ }\href@noop {} {\bibfield  {journal} {\bibinfo  {journal}
  {Phys. Lett. A}\ }\textbf {\bibinfo {volume} {163}},\ \bibinfo {pages} {239}
  (\bibinfo {year} {1992})}\BibitemShut {NoStop}%
\bibitem [{\citenamefont {Uhlmann}(1976)}]{uhlmann1976transition}%
  \BibitemOpen
  \bibfield  {author} {\bibinfo {author} {\bibfnamefont {A.}~\bibnamefont
  {Uhlmann}},\ }\href@noop {} {\bibfield  {journal} {\bibinfo  {journal} {Rep.
  Math. Phys.}\ }\textbf {\bibinfo {volume} {9}},\ \bibinfo {pages} {273}
  (\bibinfo {year} {1976})}\BibitemShut {NoStop}%
\bibitem [{\citenamefont {Jozsa}(1994)}]{jozsa1994fidelity}%
  \BibitemOpen
  \bibfield  {author} {\bibinfo {author} {\bibfnamefont {R.}~\bibnamefont
  {Jozsa}},\ }\href@noop {} {\bibfield  {journal} {\bibinfo  {journal} {J. Mod.
  Opt.}\ }\textbf {\bibinfo {volume} {41}},\ \bibinfo {pages} {2315} (\bibinfo
  {year} {1994})}\BibitemShut {NoStop}%
\end{thebibliography}
\end{document}